\documentclass[epjST]{svjour}

\usepackage{amsmath,amssymb,bbm}
\usepackage{float}
\usepackage{imakeidx}
\usepackage[colorlinks=true,linkcolor=blue,citecolor=blue,urlcolor=magenta]{hyperref}
\usepackage{footnote,comment}
\usepackage{doi}
\usepackage{enumitem}
\usepackage{slashed}
\usepackage{xcolor,xstring,xparse}
\usepackage{graphicx}
\usepackage{rotating}
\usepackage{pdflscape}
\usepackage{aastex_hack}
\usepackage{appendix}
\usepackage{booktabs}

\makesavenoteenv{tabular}
\makesavenoteenv{table}
\makesavenoteenv{figure}

\newcommand{\orcidicon}{\includegraphics[width=0.32cm]{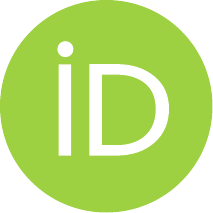}}
\newcommand{\orc}[1]{\href{https://orcid.org/#1}{\orcidicon}}

\newcommand{\orcJR}{0000-0001-8217-1484}
\newcommand{\orcAJS}{0000-0001-5474-2649}

\newcommand*{\MeV}{\text{\,MeV}}

\newcommand*{\eV}{\text{\,eV}}

\newcommand*{\beqn}{\begin{equation}}
\newcommand*{\eeqn}{\end{equation}}
\newcommand{\beql}[1]{\begin{equation} \label{#1}}
\newcommand{\eeql}[1]{\label{#1} \end{equation} } 
\newcommand{\req}[1]{Eq.\,(\ref{#1})}
\newcommand{\rf}[1]{Fig.\,{\ref{#1}}}

\newcommand{\rsec}[1]{Sec.\,{\ref{#1}}}

\newcommand{\E}{\mathrm{e}}

\NewDocumentCommand{\allcite}{m o}{\nocite{#1}\hyperlink{cite.#1}{\StrBefore{#1}{:}\IfNoValueTF{#2}{ et. al. }{ and #2 }(\StrBehind{#1}{:}[\temp]\StrLeft{\temp}{4})}}
\NewDocumentCommand{\aucite}{m o}{\nocite{#1}\hyperlink{cite.#1}{\StrBefore{#1}{:}\IfNoValueTF{#2}{ }{ and #2 }(\StrBehind{#1}{:}[\temp]\StrLeft{\temp}{4})}}
\NewDocumentCommand{\allcitep}{m o}{\nocite{#1}\hyperlink{cite.#1}{\StrBefore{#1}{:}\IfNoValueTF{#2}{ et. al. }{ and #2 }(preprint \StrBehind{#1}{:}[\temp]\StrLeft{\temp}{4})}}

\begin{document}

\title{Short Note on Spin Magnetization in QGP
}

\author{
Andrew~Steinmetz${}^1$\thanks{Author 
\email{ajsteinmetz@arizona.edu}}\orc{\orcAJS},
Johann~Rafelski${}^1$\orc{\orcJR}
}

\institute{
${}^1$Department of Physics, The University of Arizona, Tucson, AZ, 85721, USA\\
}

\abstract{
We outline the theory of spin magnetization applicable to the QGP (quark-gluon plasma) epoch of the Universe. We show that a fully spin-polarized single flavor up-quark gas could generate a cosmic magnetic fields in excess of \(10^{15}\) Tesla, far in excess of a possible upper limit to the primordial field. The complete multi component ferro-magnetized primordial fermion gas we consider consists of (five) nearly free electrically charged quarks, and leptons (electrons, muons, tau). We present details of how the magnetization is obtained using a grand partition function approach and point to the role of the nonrelativistic particle component. In the range of temperature \(150\) MeV to \(500\) MeV our results are also of interest to laboratory QGP experiments. We show that the required polarization capable to explain large scale structure magnetic fields observed has $1/T$ scaling in the limit of high $T$, and could be very small, at pico-scale. In the other limit, as temperature decreases in the expanding Universe, we show that any magnetic fields present before hadronization can be carried forward to below quark confinement condition temperature by polarization of electrons and muons.
}

\maketitle
\section{Introduction}
\label{sec:introduction}
We address the origin mystery of large scale \((\sim1\,\mathrm{Mpc})\) magnetic fields in the Universe. As it is hard to destroy magnetic fields once generated, it is possible that measured contemporary fields have primordial cosmic origins. As a first step, one can extrapolate present day observations to the primordial epoch, and then in a second step, seek appropriate formation mechanisms. Recently we have proposed the possibility that cosmic magnetism originates in the spin polarization of electron-positron pairs~\cite{Steinmetz:2023nsc,Steinmetz:2023ucp} near to the Big-Bang Nucleosynthesis epoch~\cite{Grayson:2023flr,Grayson:2024uwg}. We now approach the possible role of light quark-antiquark pairs in the QGP (quark-gluon plasma) phase near to hadronization, pushing the possible source of spin magnetization further back in time to before matter hadronization.

\begin{figure}
\centerline{
\includegraphics[width=0.90\columnwidth]{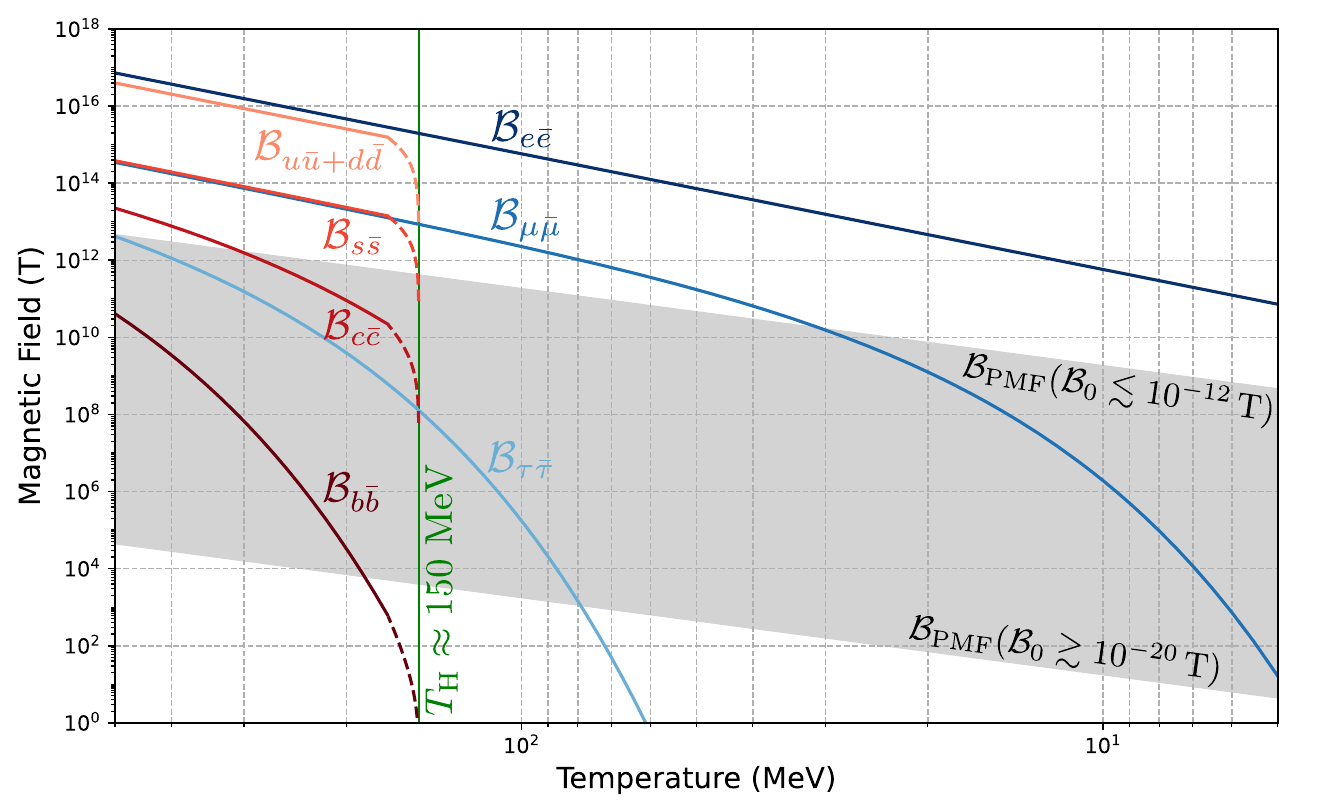}}
\caption{\label{Figure_1}Temperature dependence of several key magnetic field contributions in the early Universe during the QGP epoch. The primordial epoch magnetic fields range (grey band) was obtained from~\req{eq:PMF}. The maximum possible spin magnetization for quarks are plotted with the following curves: light \(u+d\) quarks (orange), strange (bright red), charm (dark red), and bottom (brown). Lepton curves are given by: electrons (dark blue), muons (blue), and tau (light blue). The maximum magnetic field strengths from spin polarization \(\mathcal{B}\) for charged leptons and quarks was determined by~\req{eq:HeavyNumberDensity} (summing over the first ten terms with \(k=10\)) and~\req{eq:UpperLimitMagnetization} respectively.}
\end{figure}

The existence of intergalactic fields is known from Faraday rotation measurements from distant radio active galaxy nuclei (AGN)~\cite{Pomakov:2022cem}. Conversely, measurements of synchrotron radiation from ``blazar'' AGN whose jets are pointed towards the Earth provide the lower bound on intergalactic magnetic fields~\cite{Neronov:2010gir}. Furthermore, primordial fields may also provide a solution to the ``Hubble tension'' problem in cosmology~\cite{Jedamzik:2025cax}. What makes the QGP epoch particularly attractive as the origin epoch for primordial magnetic fields (PMF) is the high matter-antimatter density involved. Therefore, the amount of spin polarization required for PMF generation is very small. This is shown in~\rf{Figure_1} over the temperature range \(500\,\MeV>T>5\,\MeV\) where \(T\) is considered in units of energy with \(k_{B}\to 1\). The PMF strength generated by maximum spin magnetization is shown for each particle species. The gray-shaded region represents the allowed PMF range, obtained by scaling today’s intergalactic magnetic field (bounded between \(10^{-12}~\mathrm{T} > \mathcal{B}_{0} > 10^{-20}~\mathrm{T}\))~\cite{Pshirkov:2015tua,Planck:2015zrl,Jedamzik:2018itu} via
\begin{equation}
\label{eq:PMF}
\mathcal{B}_\mathrm{PMF}(T) = \mathcal{B}_{0}\frac{T^{2}}{T_{0}^{2}}\,,\qquad
k_{B}T_{0} = 2.35\times10^{-4}\eV\,.
\end{equation}
which preserves magnetic flux and where \(T_{0}\) is the contemporary temperature of the CMB today.~\req{eq:PMF} is valid for under-dense regions of deep intergalactic space mostly undisturbed by contemporary magnetic fluxes generated via Amp{\`e}rian currents. For a more realistic description,~\req{eq:PMF} is multiplied by a dampening function related to the composition of the Universe~\cite{Banerjee:2004df}. Under the conditions of a deep undisturbed void, conservation of flux leads to a cosmic magnetic constant parameter given by
\begin{equation}
\label{eq:MagneticConstant}
b_{0} \equiv \frac{e\mathcal{B}(t)}{T^{2}(t)}=\frac{e\mathcal{B}_{0}}{T_{0}^{2}}\,,\qquad
10^{-3} > b_{0} > 10^{-11}\,.
\end{equation}
We note that the magnetic constant parameter may also be further dampened due to heat dissipation of magneto-hydrodynamic modes especially during neutrino decoupling and the recombination epochs~\cite{Jedamzik:1996wp}. For further discussion of~\req{eq:MagneticConstant}, see Ref.~\cite{Steinmetz:2023nsc}. The quark contribution to magnetization shown in~\rf{Figure_1} vanishes as hadronization occurs; it is assumed that the fraction of phase-space occupied by QGP evolves linearly over this period which is denoted by the dashed curves. Such an assumption is sensitive to the equations of state and the dynamics of the phase transformation~\cite{Fromerth:2012fe}. Hadronization is marked by the green vertical line at \(T_\mathrm{H}\approx150\MeV\) representing the Hagedorn temperature~\cite{Hagedorn:1967dia,Rafelski:2015xej}. The drop-off of \(\mathcal{B}_{q\bar{q}}\) occurs because of the phase transformation of QGP at \(T_\mathrm{H}\) to much less magnetically relevant heavy hadrons.

We understand the primordial deconfined QGP phase of matter due to several decades of experimental effort; this new state of matter existed in the Universe for nearly \(25~\mathrm{\mu s}\) after the Big-Bang~\cite{Rafelski:2019twp,Rafelski:2023emw,Rafelski:2024fej}. However, there are differences between the QGP produced in the early Universe versus QGP produced in laboratory heavy-ion collisions. Of greatest importance is the presence of the lepton abundance in the early Universe. Laboratory formed QGP drops are too short-lived and too small to support a comparable high-density of leptons. The net baryon content of QGP-drops formed in laboratory experiments can also be vastly different from early Universe conditions where baryon-antibaryon asymmetry is nano-scale at \(10^{-9}\); see Ref.~\cite{Fromerth:2012fe} for a discussion based on baryon chemical potential. The baryon asymmetry is characterized by the net baryon-to-photon ratio given today by \(\eta_{\gamma}=6.14\pm0.02\times10^{-10}\)~\cite{ParticleDataGroup:2022pth}. A high QGP baryon content is found at relatively low energy heavy-ion collisions near to the presumed threshold for QGP formation~\cite{Letessier:2005qe}. In typically fixed target CERN experiments (NA61 today), baryon-rich conditions are explored and also expected to be present in astrophysical compact objects~\cite{Ghosh:2025sjn}. However, as the collision energy increases towards the highest available today, the incoming nuclear valance quarks escape from the QGP drop: CERN-LHC created QGP-drops as observed by ALICE and CMS experiments have relatively low net baryon density mirroring the prevailing conditions in the primordial Universe~\cite{Letessier:2005qe}. 

In this work we expand our prior spin magnetization study~\cite{Steinmetz:2023nsc,Steinmetz:2023ucp} to consider the role of light-quark magnetization in the primordial QGP Universe, focusing on the interplay between quarks, leptons, and magnetic fields. The presence of strong magnetic fields in the primordial QGP Universe could have significantly affected the equilibrium properties of Standard Model particles in the earliest moments after the Big-Bang~\cite{Durrer:2013pga,Subramanian:2015lua}. EM response of QGP is of considerable theoretical interest~\cite{Grayson:2022asf,Shovkovy:2022bnd,Ghosh:2024fkg} and such magnetic fields have long been thought to be connected to baryon asymmetry~\cite{Vachaspati:1991nm,Baym:1995fk}. Chiral magnetism in QGP has also been studied~\cite{Fukushima:2008xe,Boyarsky:2011uy,Bali:2011qj}.

\section{Estimation of maximum plasma spin magnetization\label{sec:estimation}}
We show now that lepton- and quark-magnetism in cosmic epoch QGP can be substantial. We first obtain the upper limit of possible PMF originating in spin magnetization in the cosmic gas, and then estimate the required degree of polarization to generate contemporary PMFs. In Table~\ref{tab:particle_properties}, we list the relevant properties of select particles present in the QGP epoch of the Universe. The magnetic moment \(\mu\) is given in units of the Bohr magneton \(\mu_{B}\equiv e\hbar/2m_{e}\approx5.788\times10^{-11}\ \mathrm{MeV\;T}^{-1}\). The degrees of freedom (dof.) \(\mathfrak{g}=n_\mathrm{S}\times n_\mathrm{C}\) is the number of spin \(n_\mathrm{S}\) and color \(n_\mathrm{C}\) states available to the particle. We evaluate the magneton with gyromagnetic factor \(g=2\), while strong interaction corrections suggest a larger value for quarks. For each particle seen in Table~\ref{tab:particle_properties}, there is an antiparticle with opposite sign of magnetic moment. 
\begin{table}
\centering
\caption{Properties of select particles of relevance in primordial plasma. Relative magneton values are shown up to six decimal (though not necessarily significant) places to highlight the change in order of magnitude between particles.}
\label{tab:particle_properties}
\begin{tabular}{@{}lllll@{}}
\toprule
\textbf{Particle} & \textbf{Mass} \([\approx\mathrm{MeV}]\) & \textbf{Charge \([e]\)} & \textbf{Magneton} \([\mu/\mu_{B}]\) & \(\mathfrak{g}\) \textbf{dof.} \\ 
\midrule
Electron \((e)\) & 0.511 & \(-1\) & \(-1.001160\) & 2 \\
Muon \((\mu)\) & 105.7 & \(-1\) & \(-0.004834\) & 2 \\
Tau \((\mu)\) & 1776.9 & \(-1\) & \(-0.000286\) & 2 \\
\midrule
Up \((u)\) & 2.2 & \(+2/3\) & \(+0.154848\) & 6 \\
Down \((d)\) & 4.7 & \(-1/3\) & \(-0.036241\) & 6 \\
Strange \((s)\) & 96 & \(-1/3\) & \(-0.001793\) & 6 \\
Charm \((c)\) & 1270 & \(+2/3\) & \(+0.000267\) & 6 \\
Bottom \((b)\) & 4180 & \(-1/3\) & \(-0.000041\) & 6 \\
\bottomrule
\end{tabular}
\end{table}

The electron-positron and light-quark gases, especially up-quarks, are the magnetically most relevant particles in the QGP epoch due to their charge and low mass. The up-quark content is comparable to that of electrons since both are very relativistic particles considering \(k_{B}T=300\,\mathrm{MeV}\gg m_i c^{2}\). The number density \(n_{i}\) for the \(i\)-th light particle \((i\rightarrow u,d,e)\) is then described by a massless fermion gas given by~\cite{Letessier:2002ony}
\begin{align}
\label{eq:NumberDensity}
n_{i} = \frac{\mathfrak{g}_{i}}{2\pi^{2}}\frac{3\zeta(3)}{2}\left(\frac{k_{B}T}{\hbar c}\right)^{3}\,,
\end{align}
where \(\zeta(3)\approx1.202\) is the Riemann zeta function. The ratio of contribution to magnetism from light-quarks compared to leptons is thus solely rooted in their comparable magnetic moment and greater degeneracy. We also wish to describe the magnetic behavior of the heavier particles in the primordial plasma consisting of muons, strange quarks etc. We however omit discussion of the top-quark as it has vanished from the particle inventory of the Universe in the temperature range considered. The heavy \(i'\)-th particles \((i'\rightarrow \mu,\tau,s,c,b)\) are described by the Boltzmann expansion of the Fermi distribution function~\cite{Letessier:2002ony,Yang:2024ret}, giving a number density compared to~\req{eq:NumberDensity} of
\begin{equation}
\label{eq:HeavyNumberDensity}
n_{i'} = \frac{\mathfrak{g}_{i'}}{2\pi^{2}}\left(\frac{k_{B}T}{\hbar c}\right)^{3}\sum_{k=1}^{\infty}\frac{(-1)^{k+1}}{k^4}\left(k\frac{m_{i'}c^{2}}{k_{B}T}\right)^{2}K_{2}\left(k\frac{m_{i'}c^{2}}{k_{B}T}\right)\,,
\end{equation}
where \(K_{2}(x)\) is the modified Bessel function of the second kind (with index 2). The Boltzmann approximation is arrived at by keeping only the first \(k=1\) term in the expansion.

The estimated total cosmic magnetic flux strength is therefore derived from the sum of current-generated flux density, and the medium polarization of the most magnetically active particles \((i)\) and antiparticles \((\bar{i})\), given by
\begin{align}
\label{eq:TotalMagnetization}
\mathcal{B}_\mathrm{total}=\mathcal{B}_\mathrm{currents}+\frac{\mu_{0}}{V}\sum_{i} \mathcal{M}_{i\bar{i}}\,,
\end{align}
where \(\mathcal{M}\) is the magnetization, and \(\mu_{0}\) is the vacuum permeability (not to be confused with magnetic moment). Using~\req{eq:NumberDensity} and~\req{eq:TotalMagnetization}, we obtain an upper bound for the up-quark contribution to magnetic field strength
\begin{align}
\label{eq:UpperLimitMagnetization}
\mathcal{B}_\mathrm{max}(T) < \mu_{0}\left(\sum_{i}\mu_{i}n_{i}(T)+\sum_{\bar{i}}\mu_{\bar{i}}n_{\bar{i}}(T)\right)\,,
\end{align}

We choose to evaluate the maximum magnetization at a benchmark temperature of \(k_{B}T=300~\mathrm{MeV}\). This is because at (and above) such temperatures, QCD can be evaluated perturbatively in agreement with lattice results~\cite{ParticleDataGroup:2024cfk}. For both quarks \((q)\) and charged leptons \((\ell)\), we obtain
\begin{equation}
\label{eq:QuarkLeptonEstimate}
\mathcal{B}_{q\bar{q}}(300\MeV) < 9.1\times10^{15}~\mathrm{T} \,,\qquad
\mathcal{B}_{\ell\bar{\ell}}(300\MeV) < 1.6\times10^{16}~\mathrm{T} \,.
\end{equation}
The values presented in~\req{eq:QuarkLeptonEstimate}, based on~\req{eq:UpperLimitMagnetization}, are estimates that assumes that all strongly interacting quark and lepton magnetic moments align in a suitable manner. This is comparable to the estimated maximum stellar core magnetic field strength within magnetars~\cite{Ferrer:2010wz} and \(\sim 10^{4}\) times stronger than their estimated surface field strength~\cite{Kaspi:2017fwg}.

The electron contribution is comparable to the up-quark contribution (also shown in the electron and light-quark lines in~\rf{Figure_1}) via the ratio of degrees of freedom and the magnetic moment size
\begin{align}
\label{eq:ElectronMagnetization}
&\frac{\mathcal{B}_{e\bar{e}}}{\mathcal{B}_{u\bar{u}}} = \frac{\mathfrak{g}_{e}}{\mathfrak{g}_{u}} \frac{\mu_{e}}{\mu_{u}} \sim 2.2\,.
\end{align}
Prior to hadronization, and especially in the perturbative regime, we expect the physical spin magnetization to also agree with this ratio.

We consider~\req{eq:UpperLimitMagnetization} the maximum attainable field strength present in the primordial QGP. The actual magnetization of quarks and leptons is of course weaker due to the high temperatures of the cosmic plasma which tends to disrupt the necessary alignment required for magnetization. Figure~\ref{Figure_1} also reveals that the heaviest considered particles such as bottom or tau have densities during this era requiring a large polarization fraction required to account for the primordial fields unlike the lighter species. Heavier species behaving semi- or non-relativistically, however, may proportionally be easier to magnetize due to the minimization of free energy of the system.

We can estimate the required scaling with $T$ of the polarization factor at high temperature based on the scaling of the primordial fields in~\req{eq:PMF} and effectively massless relativistic particle abundance according to~\req{eq:NumberDensity}
\begin{equation}
\label{eq:FractionAligned}
f(T) = \frac{\mathcal{B}_\mathrm{PMF}(T)}{\mathcal{B}_{i\bar{i}}} 
= \frac{2}{3\zeta(3)}\frac{2\pi^{2}\mathcal{B}_{0}}{\mu_{0}\mu_{i}\mathfrak{g}_{i}}\left(\frac{\hbar^{3}c^{3}}{T_{0}^{2}T}\right)\propto\frac{1}{T}\,.
\end{equation}
It is then of interest to determine the minimum fraction \(f(T)\) of aligned fermions needed to account for the primordial magnetic field. If we take the ratio of the lower-bound of the PMF (see~\rf{Figure_1} and~\req{eq:PMF}) and the upper-limit magnetization in~\req{eq:UpperLimitMagnetization}, we can estimate the required fraction of aligned fermions. For illustration we evaluate \(f(T=300~\mathrm{MeV})\), yielding a range for the alignment fraction
\begin{equation}
\label{eq:FractionEstimate}
10^{-12} < f(300\MeV) < 10^{-4}\,;
\end{equation}
\req{eq:FractionEstimate} suggests that even a weakly polarized (\(10^{-12}\) pico-scale) primordial lepton-quark Fermi gas would have had a significant impact on the early Universe consistent with contemporary cosmic magnetism, as shown in~\rf{Figure_1}, with light-quarks contributing on par with leptons. While leptons remain dominant (within about a factor of 2; cf.~\req{eq:ElectronMagnetization}), they are not the sole source of Fermi spin magnetization. We provide a theoretical outline and point to where future efforts may be directed.

\section{Theory of a spin magnetized gas\label{sec:magnetization}}
Given the estimates presented in \rsec{sec:estimation}, we work towards a more realistic theory of fermion spin magnetization. We consider a free but magnetized fermion gas in the temperature range \(500\MeV>T\gtrsim150\MeV\) composed of quark and lepton particles (and antiparticles). As the magneton scales with \(\mu \propto 1/m\), the lightest species specifically have the largest magnetic moments; see Table~\ref{tab:particle_properties}. For relativistic species under the conditions of thermal and chemical equilibrium~\cite{Elze:1980er}, as was the case in the primordial Universe, the chemical potential \(\Omega\) of each particle is opposite in sign to that of its antiparticle
\begin{align}
\label{eq:equilibirum_conditions}
\Omega_{q}=T\ln\lambda_{q}\,,\qquad
\lambda_{q}=1/\lambda_{\bar{q}}\,,\qquad
\Omega_{q}=-\Omega_{\bar{q}}\,,
\end{align}
where \(\lambda\) is the fugacity. The magnetic dipole of a particle is also opposite in sign to its antiparticle \(\mu_{i}=-\mu_{\bar{i}}\) as charge is flipped. Any deviation from this condition would represent a violation of CPT symmetry~\cite{Colladay:1996iz,Bluhm:1997ci,BASE:2016yuo}.

To describe enhancement of the spin polarization of the gas caused by generalized interactions, we introduce a magnetic potential \(\Pi\) that couples to the intrinsic spin and induces polarization. Analogous to the chemical potential \(\Omega\), the magnetic polarization potential modifies the effective chemical potential for each spin state. We define the spin-dependent chemical potential as
\begin{align}
\label{eq:GeneralPotential}
\Sigma(\sigma,s) = \sigma\Omega + s\,\Pi\,,
\end{align}
so that the effective shift is \(+\Pi/2\) for \(s=+1/2\) and \(-\Pi/2\) for \(s=-1/2\). This represents potentials for each of the four components of the Dirac equations. The presence of a non-zero spin polarization potential then generates magnetization of the gas via a bias in the particle number. We briefly take a moment to comment on the physical origins of non-zero \(\Pi\). From its construction, it is most apparently associated with the magnetic dipole energy, which is demonstrated below in \rsec{sec:magnetization_evaluation}. However, we point out that magnetic energy is not the only source of polarization for species with other forms of charge. Quarks which have color charge, should also be able to participate in color magnetism. Color ferromagnetism has been speculated to be responsible for the magnetization of neutron stars~\cite{Iwazaki:2005nr,Miransky:2015ava}. We suggest that color dipoles may also be relevant in primordial QGP providing a source for polarization as color magnetic dipoles are far larger \(\sqrt{\alpha_\mathrm{S}}/m_{q}\gg\sqrt{\alpha_\mathrm{EM}}/m_{q}\).

During the QGP period, particle-antiparticle pairs of quark-antiquarks were freely produced and annihilated through photon- and gluon-mediated processes, represented by \(q+\bar{q}\rightleftharpoons2\gamma\) and \(q+\bar{q}\rightleftharpoons2g\). We note that the entropy conserving expansion of the Universe is extremely slow compared to the relevant collision reaction times during the QGP epoch~\cite{Yang:2024ret}. Accounting for the internal energy $U$ of magnetized QGP, including the energies of neutrinos~\cite{Birrell:2014ona}, involves the following properties: 
\begin{itemize}
\item[(a)] The energy of adding or removing a baryon $\Omega_{B}B$ where \(B\) is baryon number,
\item[(b)] the energy of adding or removing a lepton $\Omega_{\ell}(N_{\ell}-N_{\ell})$ where $\ell$ is any lepton, 
\item[(c)] the magnetic energy $\mathcal{M}\mathcal{B}$ where $\mathcal{M}$ is the net magnetization and $\mathcal{B}$ is the magnetic field strength and
\item[(d)] the electromagnetic energy density generated by the cosmic magnetic field.
\end{itemize}

The dependency of $U$ on $\mathcal{M}$ reflects that $\mathcal{B}$ is the incremental energy cost to change the magnetization by flipping the spin of a particle~\cite{Bali:2014kia}. Therefore, this makes magnetization $\mathcal{M}$ an extensive property of the system which changes with particle number. We see this explicitly by writing the total spin magnetization as the sum over all particles $i\in{1,\ldots,k}$
\begin{align}
\label{eq:dipole}
\mathcal{M} = \sum_{i=1}^{k}(\mu_{i}N_{i}^{\uparrow} + \mu_{\bar{i}}N_{\bar{i}}^{\uparrow} - \mu_{i}N_{i}^{\downarrow} - \mu_{\bar{i}}N_{\bar{i}}^{\downarrow})\,,\qquad
N_{i} = N_{i}^{\uparrow} + N_{i}^{\downarrow}\,.
\end{align}
The $\uparrow\downarrow$ notation refers to spin-up $(\uparrow)$ and spin-down $(\downarrow)$ states along the direction of the external field. Therefore, $N_{i}^{\uparrow\downarrow}$ refers to the $i$-th constituent population number in either spin-up or spin-down orientation. The signs of each term in~\req{eq:dipole} arises from the sign of the spin eigenvalue. While~\req{eq:dipole} includes contributions from each particle with a magnetic dipole, we expect the magnetization to be dominated by electron-positrons and the lightest quarks.
\begin{align}
\notag\mathcal{M} = &+|\mu_{u}|(N_{u}^{\uparrow}-N_{\bar{u}}^{\uparrow})-|\mu_{u}|(N_{u}^{\downarrow}-N_{\bar{u}}^{\downarrow})\\
\notag &-|\mu_{d}|(N_{d}^{\uparrow}-N_{\bar{d}}^{\uparrow})+|\mu_{d}|(N_{d}^{\downarrow}-N_{\bar{d}}^{\downarrow})\\
\label{eq:dipole2}
&-|\mu_{e}|(N_{e}^{\uparrow}-N_{\bar{e}}^{\uparrow})+|\mu_{e}|(N_{e}^{\downarrow}-N_{\bar{e}}^{\downarrow})+\ldots
\end{align}

We recognize that~\req{eq:dipole2} contains terms representing asymmetry in the spin alignment though we can organize them in two different ways: (a) we group terms of the same spin alignment or (b) we group terms of matter and antimatter. The second approach allows the definition of spin-asymmetry in terms of conserved quantities characterizing spin angular momentum. We define net spin-asymmetry numbers $\delta_{i}^{\uparrow\downarrow}$ and write
\begin{align}
\delta_{i}^{\uparrow\downarrow} &\equiv N_{i}^{\uparrow\downarrow}-N_{\bar{i}}^{\uparrow\downarrow}\,,\\
\mathcal{M} &= 
+|\mu_{u}|(\delta_{u}^{\uparrow}-\delta_{u}^{\downarrow})
-|\mu_{d}|(\delta_{d}^{\uparrow}-\delta_{d}^{\downarrow})
-|\mu_{e}|(\delta_{e}^{\uparrow}-\delta_{e}^{\downarrow})+\ldots
\end{align}
The net spin-asymmetry is the asymmetry of particles and antiparticles of the same spin. Therefore $\delta_{u}^{\uparrow}\neq0$ represents a situation where there are more up-quarks than up-antiquarks in the spin-up $\uparrow$ state.

\section{Magnetized grand partition function}
\label{sec:partition}
The partition function allows us to calculate various thermodynamic quantities by taking appropriate derivatives of the grand potential $\mathcal{F}$. The relevant contributions to the magnetized primordial plasma arise from the quarks, gluons, leptons, and the vacuum. The grand potential in terms of the grand partition function $\mathcal{Z}$ is
\begin{align}
\label{eq:parts}
\mathcal{F} &= -T\ln\mathcal{Z}\,,\\
\ln\mathcal{Z}_{\mathrm{total}} &=
\ln\mathcal{Z}_{\mathrm{quarks}} +
\ln\mathcal{Z}_{\mathrm{gluons}} +
\ln\mathcal{Z}_{\mathrm{leptons}}+
\ln\mathcal{Z}_{\mathrm{vac.}}+\ldots 
\end{align}
In the temperature range considered $(500\MeV>T\gtrsim150\MeV)$, the lightest quarks act as essentially massless particles. Heavier quarks (such as strange) can also be included albeit with mass corrections. It is worth remarking on the uniqueness of the situation: As magnetic moment scales inverse with mass, it is the particles which are most massless in character which contribute most to magnetization. The following section is written in natural units where aside of \(k_{B}\to 1\) we now also take \(\hbar\to 1,\,c\to 1\) as is customary in particle physics. We approach spin polarization from the perspective of statistical models, but also point to recent work on polarization in relativistic hydrodynamics; see Refs.~\cite{Florkowski:2024cif,Bhadury:2024whs,Becattini:2024uha,Singh:2024cub}.

\subsection{Sum of Landau states}
\label{sec:Landau}
We consider a homogeneous magnetic field domain defined along the $z$-axis with magnetic field magnitude $\mathcal{B}$. The volume $V=L^{3}$ is not necessarily infinite and defines the size of the homogeneous domain such that $\partial\mathcal{B}_{i}/\partial x_{j}\approx0$ for \(i,j=1,2,3\). For a fermion species of charge $Q$, mass $m$, and g-factor $g$, the energy eigenvalues of the magnetized particles is given by~\cite{Steinmetz:2018ryf}
\begin{align}
\label{eq:energystates}
E(p_{z},n,s)=\sqrt{m^{2}+p_{z}^{2}+2|Q|\mathcal{B}\left(n+\frac{1}{2}-\frac{g}{2}s\right)}\,,
\end{align}
where $E$ are the relativistic Landau energy eigenvalues. The micro-state energies depend on longitudinal momentum \(p_{z}\), spin $s=\pm1/2$, and Landau orbital $n=0,1,2,3,\ldots$ quantum numbers. 

The magnetized fermion partition function is then given by
\begin{equation}
\label{eq:PartitionFunction}
\ln\mathcal{Z}=\frac{n_{\mathrm{C}}V}{(2\pi)^{2}}|Q|\mathcal{B}\sum_{\sigma}^{\pm1}\sum_{s}^{\pm1}\sum_{n=0}^{\infty}\int_{-\infty}^{\infty}\mathrm{d}p_{z}\left[\ln\left(1+\exp\left(\frac{\Sigma(\sigma,s)}{T}-\frac{E(p_{z},n,s)}{T}\right)\right)\right]\,
\end{equation}
The parameter \(\sigma=\pm1\) describes both matter and antimatter states satisfying~\req{eq:equilibirum_conditions}. The Euler-Maclaurin formula is used to convert the summation over Landau levels $n$ in the integrand of~\req{eq:PartitionFunction} into an integration given by
\begin{equation}
\label{eq:EulerMaclaurin}\sum^{b}_{n=a}f(n)-\int^{b}_{a}f(x)dx = \frac{1}{2}\left(f(b)+f(a)\right)
+\sum_{i=1}^{j}\frac{b_{2i}}{(2i)!}\left(f^{(2i-1)}(b)-f^{(2i-1)}(a)\right)+R(j)\,,
\end{equation}
where $b_{2i}$ are the Bernoulli numbers and $R(j)$ is the error remainder defined by integrals over Bernoulli polynomials. The integer $j$ is chosen for the level of approximation that is desired. After some derivation,~\req{eq:PartitionFunction} can be rewritten as an integral over the three-momentum
\begin{align}
\label{eq:FreelikePartition}
\ln\mathcal{Z} &= \frac{n_\mathrm{C}V}{(2\pi)^{3}}\sum_{s}^{\pm1/2}\sum_{\sigma}^{\pm1}\int d\mathbf{p}^{3}\ln\left(1+\exp{\left(\frac{\Sigma(\sigma,s)}{T}-\frac{\sqrt{\tilde{m}^2(s)+p^{2}}}{T}\right)}\right)+\ldots\,,\\
\label{eq:spinmass}
\tilde{m}^{2}(s) &\equiv m^{2} - |Q|\mathcal{B}\,g s\,.
\end{align}
where is helpful to introduce a spin-dependent auxiliary mass \(\tilde{m}(s)\); for details see Refs.~\cite{Steinmetz:2023nsc,Steinmetz:2023ucp}. We truncate all additional terms in~\req{eq:FreelikePartition} from Euler-Maclaurin integration except the first.

We switch the partition function to spherical coordinates \(d\mathbf{p}^{3}=4\pi p^{2}dp\) and integrate by parts yielding
\begin{align}
\label{eq:partition_byparts}
\ln\mathcal{Z} &= \frac{n_\mathrm{C}V}{6\pi^{2}} \sum_{s}^{\pm1/2}\sum_{\sigma}^{\pm1}\int_{0}^{\infty} \frac{dp}{T} \, \frac{p^4}{\sqrt{\tilde{m}^2(s)+p^{2}}}F\left(\frac{\sqrt{\tilde{m}^2(s)+p^{2}}}{T} - \frac{\Sigma(\sigma,s)}{T}\right)\,,\\
\label{eq:FermiDirac}
F\left(x\right) &= \frac{1}{\exp{(x)} + 1}\,.
\end{align}
where \(F(x)\) is the Fermi-Dirac distribution. The form of the partition function expressed by~\req{eq:partition_byparts} more directly lets us evaluate thermodynamic quantities in terms of Fermi integrals~\cite{Elze:1980er,Birrell:2024bdb}. However, integrating over momentum is not an ideal description as relativistic expansions in momentum yield series that are only semi-convergent. 

To simplify the integration process, we introduce dimensionless variables by normalizing relevant physical quantities with the temperature \( T \). This approach renders the equations dimensionless and highlights the thermal contributions explicitly. The dimensionless variables are defined as
\begin{align}
\label{eq:dimensionless_variables}
p_{T} \equiv \frac{p}{T}\,, \qquad \tilde{m}_{T}(s) \equiv \frac{\tilde{m}(s)}{T}\,, \qquad \Sigma_{T}(\sigma,s) \equiv \frac{\Sigma(\sigma,s)}{T}\,.
\end{align}
This yields momentum-like \(p_{T}\), chemical potential-like \(\Sigma_{T}\) and mass-like \(\tilde{m}_{T}\) parameters. Using the relativistic dispersion relation, the dimensionless energy \(E_{T}\) can be expressed in terms of \(p_{T}\) and \(\tilde{m}_{T}\)
\begin{align}
E_{T}(p_{T},s) \equiv \sqrt{p_{T}^{2} + \tilde{m}_{T}^{2}(s)}\,.
\end{align}
The differential \( dp_{T} \) and \( dE_{T} \) transform as
\begin{align}
dp = T \, dp_{T}\,,\qquad p_{T}dp_{T} = E_{T} dE_{T}\,,
\end{align}
and the limits of integration change accordingly
\begin{equation}
p_{T} \to 0 \quad \Rightarrow \quad E_{T} \to \tilde{m}_{T}, \quad p_{T} \to \infty \quad \Rightarrow \quad E_{T} \to \infty.
\end{equation}
Substituting these dimensionless variables and differentials into the partition function \( \ln\mathcal{Z} \), we obtain expressions for both momentum-like \(p_{T}\) integration and energy-like \(E_{T}\) integration
\begin{align}
\label{eq:dimensionless_partition}
\ln\mathcal{Z} 
&= \frac{n_\mathrm{C} V}{2\pi^{2}} \frac{T^{3}}{3} \sum_{s}^{\pm1/2}\sum_{\sigma}^{\pm1} \int_{0}^{\infty} dp_{T} \, \frac{p_{T}^{4}}{\sqrt{p_{T}^{2} + \tilde{m}_{T}^{2}(s)}} \, F\left(\sqrt{p_{T}^{2} + \tilde{m}_{T}^{2}(s)} - \Sigma_{T}(\sigma,s)\right)\,,\\
\label{eq:dimensionless_partition2}
&= \frac{n_\mathrm{C} V}{2\pi^{2}} \frac{T^{3}}{3} \sum_{s}^{\pm1/2}\sum_{\sigma}^{\pm1} \int_{\tilde{m}_{T}(s)}^{\infty} dE_{T} \, (E_{T}^2-\tilde{m}_{T}^{2}(s))^{3/2} F\left(E_{T} - \Sigma_{T}(\sigma,s)\right)\,.
\end{align}
In this formulation, it is evident that the logarithm of the partition function scales as \( \ln\mathcal{Z} \propto T^{3} \), consistent with the expected thermodynamic behavior for a relativistic gas in three spatial dimensions.

\subsection{Evaluation of magnetization in the massless limit}
\label{sec:magnetization_evaluation}
The power and utility of the partition function in statistical systems is found by examining the Fermi integral in different limits and expansions. Taking the derivative of the free energy \(\mathcal{F} = -T \ln\mathcal{Z}\) with respect to the magnetic field \(\mathcal{B}\), we obtain the magnetization
\begin{align}
\label{eq:magnetization_def}
\mathcal{M} = \left( \frac{\partial \mathcal{F}}{\partial \mathcal{B}} \right) = -T \left( \frac{\partial \ln\mathcal{Z}}{\partial \mathcal{B}} \right)\,.
\end{align} 
While we emphasize that the dimensionless mass \(\tilde{m}_{T}(s)\) depends on the magnetic field and spin via~\req{eq:spinmass}, we calculate the magnetization explicitly, starting from the massless limit \(m\to 0\) with the chemical potential \(\Omega\to0\). In this limit, the dimensionless spin-dependent chemical potential simplifies to
\begin{align}
\Sigma(\sigma,s)\vert_{\Omega=0} = s\,\Pi\,,
\end{align}
where we have packaged all magnetic dependency on the polarization potential \(\Pi=\Pi(\mathcal{B})\) as a perturbation
\begin{equation}
\sqrt{p^{2}+\tilde{m}^{2}(s)}-\Sigma\approx\sqrt{p^{2}+m^{2}}-s\Pi\,.
\end{equation}
Thus, the dimensionless partition function (cf.~\req{eq:dimensionless_partition}) becomes
\begin{align}
\label{eq:MasslessPartition}
\ln\mathcal{Z}\vert_{m=0}^{\Omega=0} 
= \frac{n_\mathrm{C} V}{\pi^{2}} \frac{T^{3}}{3} \sum_{s}^{\pm1/2}\int_{0}^{\infty} dp_{T} \, p_{T}^{3} \, F\left(p_{T} - s\Pi_{T}\right)\,,
\end{align}
where \(\Pi_{T}\equiv\Pi/T\) is the dimensionless polarization potential. We note the summation over matter-antimatter states (\(\sigma=\pm1\)) yields a factor of two as no dependence on \(\sigma\) remains.

Since the polarization potential \(\Pi\) depends on the magnetic field, using~\req{eq:magnetization_def} and~\req{eq:MasslessPartition} we obtain
\begin{align}
\label{eq:MasslessMagnetization}
\mathcal{M}\vert_{m=0}^{\Omega=0} &= \frac{n_\mathrm{C}VT^{3}}{\pi^{2}}\frac{\partial\Pi}{\partial\mathcal{B}}\sum_{s}^{\pm 1/2}s\int_{0}^{\infty}dp_{T}\,p_{T}^{2}F\left(p_{T}-s\Pi_{T}\right)\,.
\end{align}
This is the exact analytical expression for the massless particle magnetization induced by a spin polarization potential \(\Pi\). If the spin potential \(\Pi\) vanishes, or becomes independent of the external magnetic field, the magnetization vanishes as expected.

The particle number \(N^{\uparrow\downarrow}\) can also be established from the partition function~\req{eq:MasslessPartition}, allowing us to write
\begin{equation}
N^{\uparrow\downarrow}\vert_{m=0}^{\Omega=0} = \pm\frac{\partial\ln\mathcal{Z}\vert_{m=0}^{\Omega=0}}{\partial(\Pi/T)}\,,\qquad
\mathcal{M}\vert_{m=0}^{\Omega=0} = \frac{\partial\Pi}{\partial\mathcal{B}}\left(N^{\uparrow}-N^{\downarrow}\right)\,,
\end{equation}
in agreement with~\req{eq:dipole} in \rsec{sec:magnetization}. This also lets us identify \(\mu\leftrightarrow\partial\Pi/\partial\mathcal{B}\) as the relationship between the magnetic dipole moment of a particle and the magnetic potential in the massless approximation.~\req{eq:MasslessPartition} and~\req{eq:MasslessMagnetization} can also be expressed in terms of polylogarithm functions.

To verify consistency, we consider the small-field (\(\Pi/T\ll 1\)) limit which corresponds to the high temperature case. Expanding the Fermi-Dirac integral for a small argument, we obtain
\begin{align}
\lim_{z\to0}\int_{0}^{\infty} dx\,\frac{x^{2}}{\exp\left(x-z\right)+1} \approx \frac{3}{4}\zeta(3)+\frac{\pi^2}{24}z\,,
\end{align}
yielding
\begin{align}
\label{eq:DensityLimit}
N^{\uparrow\downarrow}\vert_{m=0}^{\Omega=0} &\approx \frac{n_{C}V}{2\pi^{2}}\frac{3\zeta(3)}{2}T^3 \pm \frac{n_{\mathrm{C}}V}{24}\frac{\Pi}{T}T^3\,.
\end{align}
The first term in~\req{eq:DensityLimit} is the standard thermal particle number expression for a massless gas in agreement with~\req{eq:NumberDensity}, while the second term is the number shift due to polarization. The two terms also differ by one power of temperature, which verifies our earlier derivation that the polarization fraction~\req{eq:FractionAligned} goes with \(f(T)\propto1/T\).

As the standard thermal term is insensitive to polarization, it therefore vanishes when evaluating the difference \(N^{\uparrow}-N^{\downarrow}\) which appears in the magnetization, yielding
\begin{align}
\label{eq:FinalMag}
\mathcal{M}\vert_{m=0}^{\Omega=0} \approx \frac{n_\mathrm{C}V}{24}T^2\frac{\partial\Pi^2}{\partial\mathcal{B}}\,.
\end{align}
Notably, the magnetization is insensitive to the sign of \(\Pi\). In much how we expected the free energy to be \(\ln\mathcal{Z}\sim T^{3}\), we see the magnetization is \(\mathcal{M}\sim T^{2}\) in natural units via dimensional analysis. This is in agreement to our prior work~\cite{Steinmetz:2023nsc,Steinmetz:2023ucp} where we evaluated the magnetization in the Boltzmann limit~\cite{Steinmetz:2023nsc} with \(T\ll m_e\). The benefit of expressing the magnetization in the form of~\req{eq:dimensionless_partition} or~\req{eq:dimensionless_partition2} is that the integrand entirely contains the Fermi-Dirac distribution scaled by energy without mass (or magnetic fields) except as a boundary condition on the integration. This makes it suitable for numerical evaluation and comparison to the Boltzmann limit which will be the subject of future efforts.

\section{Conclusions}
\label{sec:conclusions}
We propose an alternative source of magnetism in the Universe based on primordial spin polarization. We extended here prior work~\cite{Steinmetz:2023nsc}, where we considered a plasma of electrons and positrons, to the case of primordial quark-gluon plasma. We find, see~\req{eq:FractionEstimate}, that a $10^{-12}$ polarization of up-quark and electron components would lead to cosmic primordial magnetic fields consistent with the present day intergalactic range \(10^{-12}~\mathrm{T} > \mathcal{B}_{0} > 10^{-20}~\mathrm{T}\). This suggests that a minute primordial QGP spin alignment could be an efficient driver of cosmic magnetism.

Electrons, muons, up-quarks, down-quarks, and strange-quarks all contribute to the magnetization, as illustrated in~\rf{Figure_1}. We estimated that leptonic contributions are dominant by a factor of roughly two compared to light quarks (cf.~\req{eq:ElectronMagnetization}). Strangeness and muons during QGP are less relativistic thus making them easier to magnetize yet still have sufficiently large densities to be relevant. Even heavier particle species, such as tau or charm, could also be of relevance but require  almost completely polarization which necessitates a large spin polarization potential (cf. \req{eq:GeneralPotential}). Our derivations, based on the grand partition function and direct evaluation  in the massless limit~\req{eq:FinalMag}, reveal that the magnetization scales as \(\mathcal{M}\vert_{m=0}^{\Omega=0} \approx T^{2}\) while the required alignment fraction exhibits an inverse dependence on temperature \(1/T\). This provides a starting point for future more precise study of primordial spin alignment and color ferromagnetism in systematic many body approach.

Magnetic field is sourced by electrical charge. However, the strong interactions between quarks in QGP could provide a natural mechanism for the creation of the required tiny polarization. We recall the mass difference between the nucleon and the \(\Delta\)-resonance, about \(300\MeV\), is attributed to the color hyperfine interaction of the attractive spin-spin interaction in the color sector~\cite{Johnson:1975zp,DeGrand:1975cf}. This strongly suggests that color ferromagnetism is capable near to hadronization of QGP to generate the required polarization. The maximum fields generated by a completely polarized quark gas during the QGP epoch at \(300\MeV\) would be on the order of \(10^{15}\) Tesla. This exceeds both the critical Schwinger field and the surface fields of magnetars by several orders of magnitude, and are comparable to estimates of magnetar stellar cores. Future work could explore strong interaction color ferromagnetism in QGP (of both light and heavy quarks) in more detail allowing evaluation of the overall EM magnetization of primordial QGP. 
\\

\noindent\textbf{Acknowledgment:} 
One of us (JR) thanks Tamas Bir\'o and the HUN-REN Wigner Research Centre for Physics for their kind hospitality in Budapest during the PP2024 conference, supported by the Hungarian National Research, Development and Innovation Office (NKFIH) under awards 2022-2.1.1-NL-2022-00002 and 2020-2.1.1-ED-2024-00314. This meeting and the related research report motivate the presentation of these recently obtained results. The authors and this work were not supported by any sponsor.\\
\textbf{Author Contributions:} All authors participated in every stage of the development of this work.\\
\textbf{Data Availability:} No datasets were generated or analyzed during the current study.\\
\textbf{Competing Interests:} The authors declare no competing interests.

\bibliographystyle{sn-aps}
\bibliography{short-note-qcd.bib}
\end{document}